\documentclass[a4paper]{spie}  
\usepackage[]{graphicx}
\usepackage{astro_bib_macro}

\renewcommand{\subsubsection}[1]{\noindent\textbf{#1}\newline}   

\title{PIONIER: a status report} 

\author{ J.-B. Le Bouquin\supit{a} , J.-P., Berger\supit{a,b} , G. Zins\supit{a} ,
  B. Lazareff\supit{a} , 
  L. Jocou\supit{a} , P. Kern\supit{a} , R. Millan-Gabet\supit{c} ,
  W. Traub\supit{d} , P. Haguenauer\supit{b} , O. Absil\supit{e} ,
  J.-C. Augereau\supit{a} , M. Benisty\supit{a} ,
  N. Blind\supit{a} ,
  A. Delboulbe\supit{a} , P. Feautrier\supit{a} ,
  M. Germain\supit{a} , D. Gillier\supit{a} , P. Gitton\supit{b}
  , M. Kiekebusch\supit{b} , J. Knudstrup\supit{b} , J.-L
  Lizon\supit{b} ,
  Y. Magnard\supit{a} , F. Malbet\supit{a} , D. Maurel\supit{a}
  , F. Menard\supit{a} , M. Micallef\supit{a} ,
  L. Michaud\supit{a} , T. Moulin\supit{a} ,
  D. Popovic\supit{b} , K. Perraut\supit{a} , P. Rabou\supit{a} , S. Rochat\supit{a} ,
  F. Roussel\supit{a} , A. Roux\supit{a} , E. Stadler\supit{a} and
  E. Tatulli\supit{a}
\skiplinehalf
\supit{a} UJF-Grenoble 1 / CNRS-INSU, Institut de Plan{\'e}tologie et d'Astrophysique de Grenoble (IPAG) UMR 5274, Grenoble, France\\
\supit{b} European Southern Observatory, Paranal, Chile, Garching, Germany\\
\supit{c} NextSci, California Institute of Technology, Pasadena, California USA\\
\supit{d} Jet Propulsion Laboratory, California Institute of
Technologie, Pasadena, California, USA \\
\supit{e} Universite de Li\`ege, Li\`ege, Belgium \\
}

\authorinfo{Send correspondence to J.B Le Bouquin: E-mail: lebouquj@ujf-grenoble.fr, Phone: +33 4 76 63 58 93}

\begin{document} 
\maketitle 

\begin{abstract}
The visitor instrument PIONIER provides VLTI with improved imaging capabilities and sensitivity. The instrument started routinely delivering scientific data in November 2010, that is less than 12 months after being approved by the ESO Science and Technical Committee. We recall the challenges that had to be tackled to design, built and commission PIONIER. We summarize the typical performances and some astrophysical results obtained so far. We conclude this paper by summarizing lessons learned.
\end{abstract}


\section{INTRODUCTION}
\label{sec:instrument}

The recent years may be remembered as the ``imaging revolution'' in optical interferometry with the first regular imaging results obtained at CHARA and  VLTI. By simultaneously combining four or six 1m-telescopes, the MIRC combiner at CHARA is arguably the most advanced imaging facility and the only one to provide snapshot capabilities. This was demonstrated by the impressive images of stellar surfaces by Monnier et al.\cite{monnier:2007jul} or Kloppenborg et al\cite{Kloppenborg:2010}.

Considering the available baseline lengths in modern interferometers, Monnier et al. conclude that sensitivity is not an issue.\footnote{Monnier: ``In order for imaging to be useful on a complicated object, the characteristic size scale must be many angular resolution elements across. For broadband imaging of stars, this ``size'' requirement assures us that suitable imaging targets are always ``bright'' by even interferometric standards. [...] In a world with kilometric baseline interferometers, sensitivity will again become an issue for imaging more distant stars.''} However this is mainly true when imaging stellar surfaces because of their high surface brightnesses compared to many other astronomical objects ($\mathrm{T}>2\,000$K, emissivity of one). The need for photons becomes way more critical when imaging emission from cooler and/or diffuse material. This includes the circum-stellar shells of mass-losing stars or the dusty disks around young stars, where planets eventually form.

With the goal to achieve \emph{sensitive imaging}, the VLTI is the most promising place. It offers the unique combination of four Unit Telescopes of $8\,$m (UTs) and four relocatable Auxiliary Telescopes of $1.8\,$m (ATs).  However, the first generation of facility instruments that will combine four telescopes simultaneously is not expected to be operational prior to 2014. Therefore, IPAG and its partners have proposed to bridge this gap with a visitor instrument combining imaging, sensitivity, and precision. Initially proposed for the period 2010-2012, the principle of PIONIER was approved by the ESO Science and Technical Committee in November 2009. The instrument was integrated at IPAG and commissioned at the Paranal Observatory in October 2010. The continuation of PIONIER operation at Paranal is discussed and renewed on a yearly basis. Current agreement lasts until October 2013.

\section{BUILDING A FAST TRACK INSTRUMENT}
\label{sec:instrument}

\subsection{Brief description}
PIONIER was first described at the 2010 SPIE meeting by JP~Berger et al\cite{Berger:2010} (initiator and first PI of the project). A description of the commissioning, including data reduction algorithms, can be found in A\&A\cite{Le-Bouquin:2011}. A nice overview of PIONIER as of 2012 is published in the ESO messenger by Zins et al\cite{Zins:2011}. PIONIER is a pairwise, temporally scanned beam combiner allowing simultaneous measurements of all available baselines and  closure phases in the 4-telescope VLTI array, either with the four 1.8m Auxiliary Telescope or the four 8m Unit Telescopes.\footnote{Recent technical tests demonstrated the feasibility of heterogenous configuration. Consequently the VLTI can virtually recombines simultaneously six and up to eight telescopes, but this requires important modifications of software and hardware that are not contemplated on a medium term basis.}

\begin{figure}[t]
   \centering
   \includegraphics[width=\textwidth]{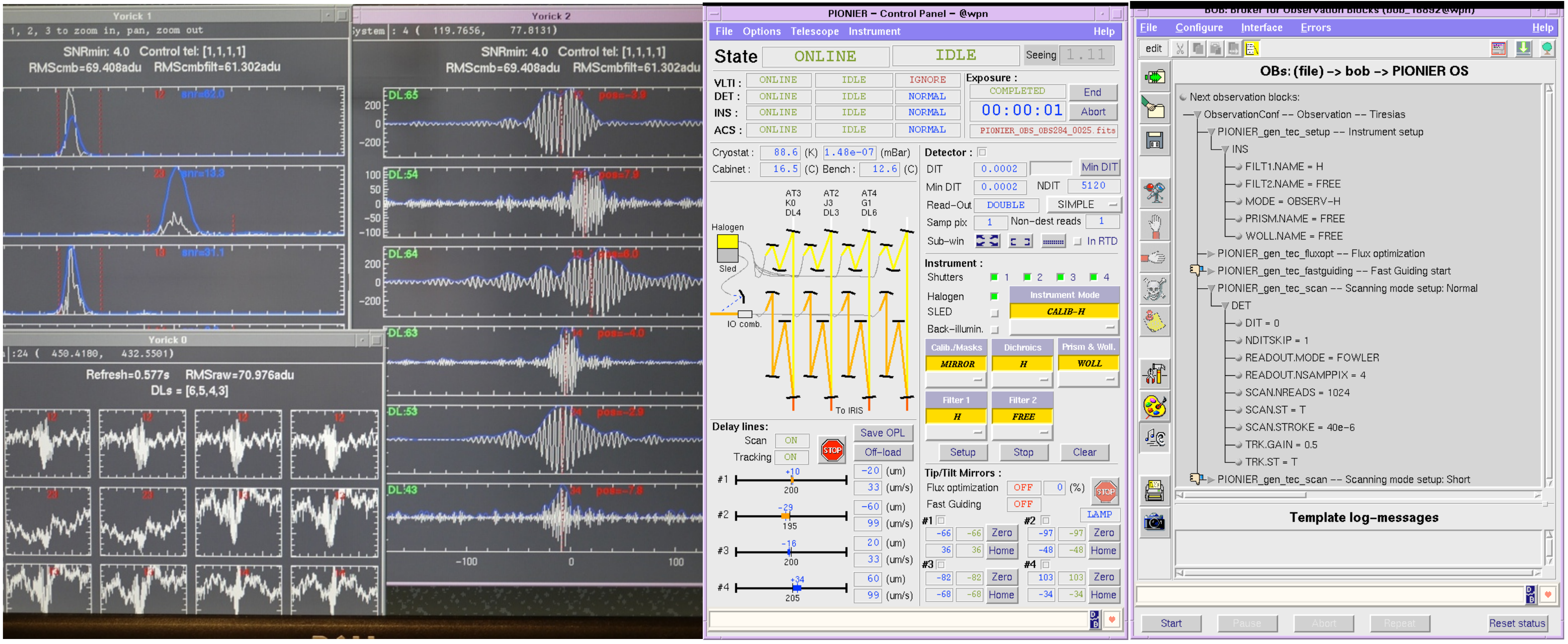} 
 \caption{A sample of the various panels used to operate PIONIER. The Real Time Display panels (left) show the raw interferograms, their power spectral densities and a filtered version of the signal. These panels also implements the real-time tracking of the fringe enveloppe. The Operating Software panel (middle) shows the status of the various subsystems and a sketch of the optical layout currently in use. The Broker of Observing Block (right) fetches and executes the OBs.}
   \label{fig:instrument}
 \end{figure}

Here we briefly summarize the main properties on the PIONIER instrument as built:
\begin{itemize}
\item Spectral range 1.45-2.5 microns (H or K bands). To date, we have only used the H-band for science observations. Switching from one band to the other requires to manually exchange opto-mechanical parts of the instrument and takes about 4h.
\item Broad band ($R\sim5$) or low-resolution spectroscopy ($R\sim15$ or $R\sim35$).  In broad band, a Wollaston prism can be inserted to simultaneously record fringes in vertical and horizontal polarisation states (laboratory axes). These functions are now motorized and can be remotely selected.
\item Spatial filtering using single mode, polarization maintaining fibers. Each beam combiner (see next item) is glued with its own four fibers. Fibers are equalized at a precision of about $0.5$mm, resulting into a negligible residual chromatic dispersion, but a large polarization dispersion ($\approx1$ fringe). This polarisation dispersion is compensated with a dedicated hardware as explained by Lazareff et al. in this proceeding.
\item Integrated Optics beam combiner (IOBC) encoding the six baselines in a pairwise scheme (see Benisty et al. \cite{benisty:2009may}). Three chips are currently available: one H-band combiner with 12 outputs (AC per baseline), one H-band with 24 outputs (ABCD per baseline) and one K-band with 24 outputs (ABCD per baseline).
\item No fringe-tracker, fringes are recorded ``on-the-fly'' thanks to a fast scanning piezo. PIONIER computes and corrects the average position of the fringe envelope at about 1 Hz (coherencing mode). Corrections are currently sent to the internal scanning piezo, although it should be possible to send them directly to the main Delay Lines of VLTI.
\item As much as possible, PIONIER complies with the VLTI standards for hardware, software and operational procedures. Figure~\ref{fig:instrument} shows the master control panels as an example of the user interfaces and of the real interferometric signal encoded by the instrument.
\end{itemize}

\subsection{Project time-line and key steps}
\label{sec:timeline}
We believe it is of general interest to detail the timeline of the PIONIER fast-track project, including some historical background and the unavoidable contingencies when building the instrument.

\begin{figure}[t]
   \centering
   \includegraphics[width=0.7\textwidth]{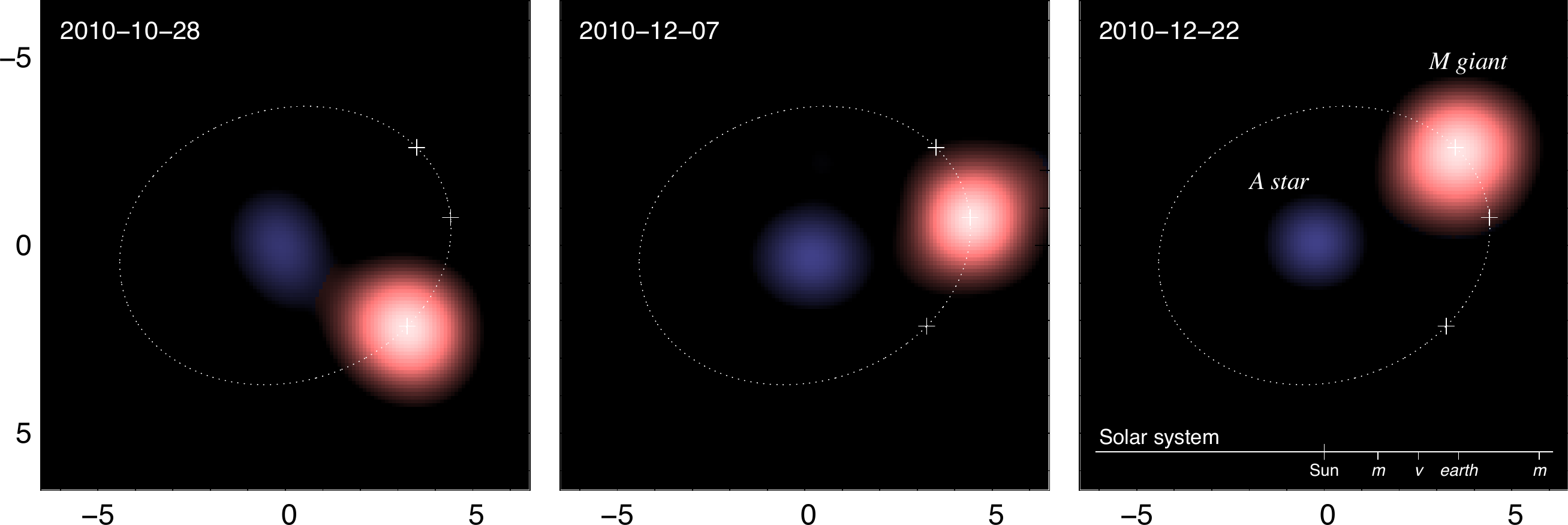} 
   \includegraphics[width=0.29\textwidth]{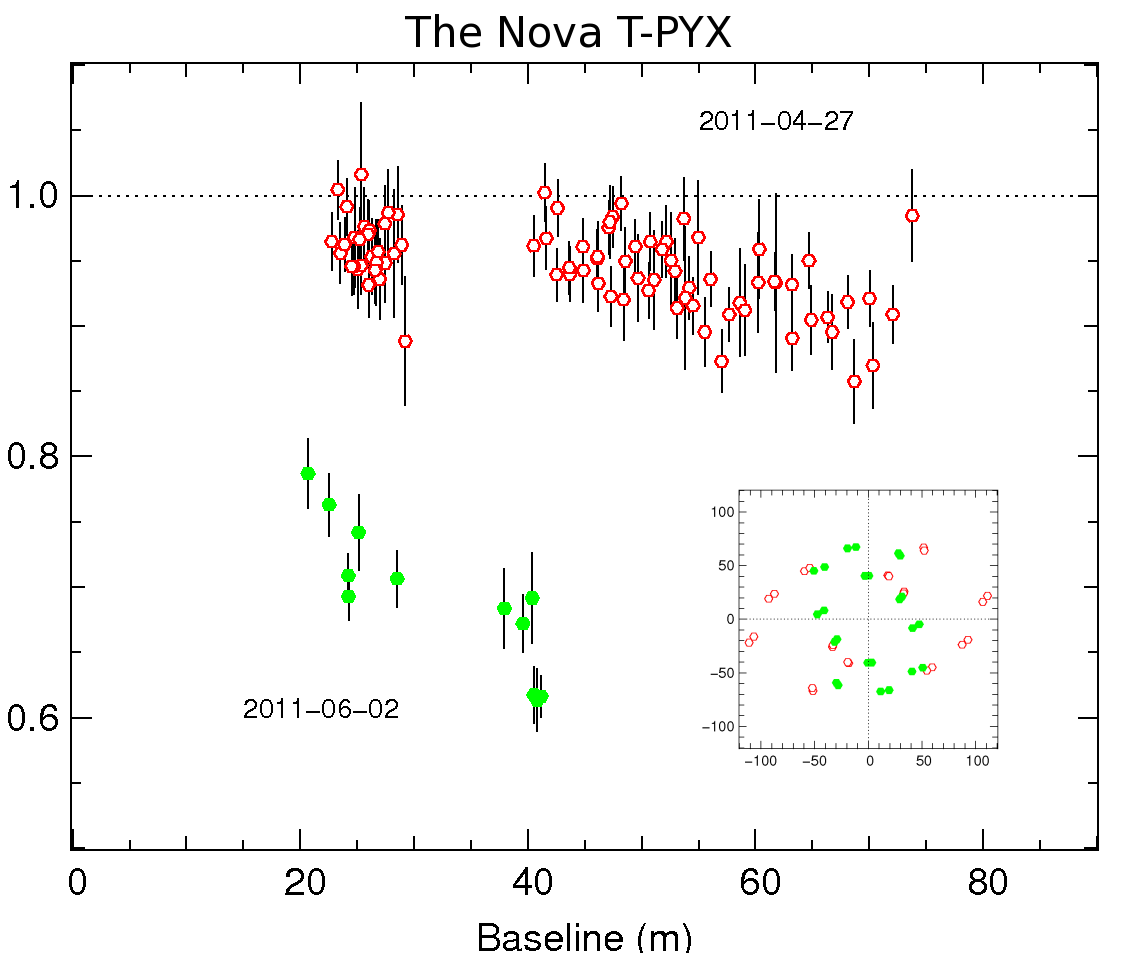} 
 \caption{First astrophysical results of PIONIER. Left: images of the symbiotic binary SS Lep captured at different orbital phases. Scale is in mas. Right: square visibility measurements of the recurrent novae T Pyx at the beginning (red) and at the end (green) of the last outburst, showing a large increase in apparent size but no departure from axial-symmetry.}
   \label{fig:first}
 \end{figure}

\begin{description}
\item[\it 2002-11] Proposition to ESO to implement a 4T extension of VINCI based on integrated technology. Although this project never started, it conceptual description was already very similar to the final PIONIER design (H-band, pairwise 4T, fast scanning piezo).
\item[\it 2008-01] Manufacturing and characterization of the first 4-beam integrated beam combiners as the result of a collaboration between IPAG and CEA-LETI in the context of the VSI project.
\item[2009-09] First funding obtained from University Joseph Fourier (Grenoble) for a team-instrument for VLTI. PIONIER is officially proposed to ESO Science and Technical Council. It is accepted as a visitor-instrument for VLTI, for the period 2010-2012.
\item[2010-01] Integration start at IPAG. First 2-beam fringes obtained in February. The large amount of polarisation dispersion introduced by the fibers is discovered in March. A working solution based on compensation with tilted lithium-niobate plates is quickly identified, and implemented in July.
\item[2010-07] First 4-beam fringes. Performance optimisation in laboratory (readout speed, opto-mechanical stability, software and communication stability). First version of the data reduction software (\texttt{pndrs}). 
\item[2010-08] Critical intervention in the detector electronic to improve the signal quality. A full differential analog signal transmission is implemented between pre-amplifier and the (new) analog to digital converter. Common mode noise rejection improves by 30db. In addition, a slow variation of the pixel offset (correlated with the detector Read state) is fixed by measuring the signal with a high impedance input. Such a drift would have prevented the use of non-destructive reads which is the main mode for PIONIER.
\item[2010-09] Accidental power outage at IPAG during several days. The PIONIER team performs the last checks before packaging running with an emergency external generator. As a positive side effect, the team stayed well focused on the duties as the e-mails were unavailable!
\item[2010-10] Shipping to Chile. Most of the opto-mechanical parts of the instrument could be shipped assembled, saving important time of alignement in Paranal. The instrument is re-assembled in Paranal and aligned with VLTI within three days. First commissioning nights: check of VLTI functionalities with four telescopes, optimisation of the operation, writing of observation templates, on-sky validation of the \texttt{pndrs}, science demonstration.
\item[2010-11] Second and last commissioning: performances check, freezing of observational procedures and templates, science verification. First science runs start.
\item[2011-03] The 4T-ABCD-H beam combiner with 24 outputs is replaced by a 4T-AC-H combiner with 12 outputs. This reduces the contribution of the detector noise and improves the scanning rate. Remember that PIONIER uses temporally scanned interferograms and therefore does not formally require the four interference states provided by the ABCD chip (see next section).
\item[2011-04] First interferometric combination of the four UTs. Useless fringes due to bad weather (almost impossible to lock the MACAO Adaptive Optics).
\item[2011-11] Daytime intervention to motorize the translation stage to allow remote switch between the broad-band and spectroscopic modes. This improvement was contemplated since the beginning of the project but could not be implemented in time before shipping.
\item[2011-11] Publication of first refereed papers based on PIONIER observations.\cite{Le-Bouquin:2011} \cite{Chesneau:2011}\cite{Blind:2011a} First results presented at the VLTI anniversary conference in Garching. Among them are the images of the symbiotic binary SS~Lep captured at different orbital phases and the expansion of the recurrent novae T~Pyx during its last outburst (see Fig.~\ref{fig:first}).
\item[2012-01] One of the fibers feeding the 4T-AC-H integrated optics component broke. A convincing explanation for the accident is still missing (no humain intervention at that time). Operation fall back to the 4T-ABCD-H, which was still available in Paranal. The 4T-AC-H component is shipped back to IPAG to be equipped with new fibers.
\item[2012-04] Upgrade of the instrument to permit K-band operation. The optics that images the outputs of the component onto the detector is upgraded to allow simpler (but still manual) switch between H-band and K-band optics. A 4T-ABCD-K integrated optics combiner is tested. Fringes are recorded on targets up to $K\sim5.5$. The main limitation is the thermal background noise, the instrument being entirely at room temperature except the detector itself. PIONIER is switched back to H-band at the end of the commissioning run. Switching between H-band and K-band requires about 4h.
\item[2012-05] First science with the four UTs. Calibrated fringes are recorded on a $H\approx11$ unresolved target, yielding $10\%$ precision squared visibilities and $5\deg$ closure phases.
\item[\it 2012-06] The 4T-AC-H component should be put back in place of the 4T-ABCD-H.
\item[\it 2013-09] Official date for the dismounting of PIONIER from the VLTI laboratory.
\end{description}

\subsection{Shortcuts imposed by the fast timing of the project}

\subsubsection{Scanning versus ABCD modulation} One can argue that the 4T-ABCD-H integrated optics component already allows an optimal measurement of the fringes with the ABCD (static) modulation only. This is entirely true. However it was decided in the early stages of the project to start with an additional temporal modulation, because of the experience of the group with this method (real-time fringe coherencing, data processing). Switching to static ABCD modulation is still easily possible from the hardware and operational software point of view. It would require a rewriting of the coherencing algorithm and of the data reduction software and dedicated commissioning time.

\subsubsection{Detector characterization} PIONIER is equipped with the former PICNIC Rockwell camera from the IOTA interferometer, kindly on loan from JPL.\footnote{Anyway it would have been impossible to buy a dedicated camera given the global budget of the project.} The custom electronics was upgraded to allow faster analog to digital conversion ($0.25\mu$sec instead of $10\mu$s conversion time). A second intervention was necessary to fix signal quality issues (see sec.~\ref{sec:timeline}). After optimizing the clocking scheme, it provides a sufficiently high frame rate ($\approx4$kHz for a single read of 12x3 pixels), with a detector noise kept below $18e^-$/read. However several behaviors of the camera could not be properly understood (or not understood at all) during the few months dedicated to the instrument integration. This includes the existence of a large transient lasting several tens of ms after the reset, or the episodic apparence of a pattern in the dark level apparently linked to an excess noise.

\section{Observing with PIONIER at VLTI}
The operation and maintenance of the instrument at Paranal are under the responsibility of the PIONIER team (during a PIONIER observing run, the operation of VLTI itself remains to the ESO Telescope and Instrument Operator). As an in-kind contribution, the Paranal observatory performs the daily refill of the detector dewar with nitrogen and routinely check the vacuum quality.

\subsection{Instrument preparation for the night}

As with any single mode instrument, care should be taken to keep PIONIER properly aligned to ensure a maximum efficiency. However the use of an integrated beam combiner considerably reduces the number of degree of freedom. At the entrance of the instrument, the injection into the single mode fibers is optimized on a daily basis thanks to motorized tip-tilt piezo. The procedure is applied  during the evening twilight, using the beacon sources located on the VLTI telescopes. On the UTs the optimization may be done on the star provided the Adaptive Optics is locking well. However, the two following adjustments should be checked \emph{manually} at the beginning of each observing run:
\begin{itemize}
\item the vertical/horizontal positioning of the IOBC, in order to optimize the coupling of the output spots into the detector pixels;
\item the angle of the lithium-niobate plates that are used to compensate for the polarization dispersion. This polarization compensation is explained in detail by Lazareff et al in this proceeding.
\end{itemize}
Although this is not contemplated, it is interesting to note that motorizing these relatively simple degrees of freedom (as well as the daily refill of the camera dewar) would permit fully remote operations of the instrument, as for the VEGA combiner at CHARA for instance.

\subsection{Observing sequence and timing} Speed and reliability of operation at VLTI considerably improved in the past few years. With PIONIER, the observation of a science target and the associated calibration star takes between 12min (bright star case with known offsets) and 30min (when including long OPD search on a faint star for instance). Consequently it is possible to gather between 15 to 35 calibrated points per night. We illustrate the operation by describing a typical observing sequence with the four ATs of VLTI.
\begin{itemize}
\item Telescope slewing: $\approx5$s to $\approx30$s.
\item Telescope acquisition on STRAP fast tip/tilt guider: $\approx3$min. This can be reduced to less than $10$s if the V-magnitude is within $\pm1$mag of the previous star because the STRAP optimisation can be skipped. A way to improve when quickly swapping between two nearby stars of different Vmag would be to saved/restore the STRAP parameters instead of optimizing each time.
\item Pupil alignment: $\approx3$min, but this should only be repeated few times over the night.
\item Acquisition on the IRIS near-IR guiding camera in the lab: $\approx1$min, dominated by the time to take a new background (less than $5$s if the same readout speed can be used so that background can be skipped).
\item Fringe search: $\approx10$s if OPD offset are already known on this target (PIONIER Operating Software save/restore offsets). $\approx2$min if OPD offsets have been found in a star close in the sky. More than $\approx5$min if OPD offsets are unknown, as one may have to scan over several millimeters. Consequently, a lot is to be gained by better OPD models, but mainly when changing target on sky. However the gain would be marginal when in a sequence SCI-CAL-SCI-CAL... because offsets of individual stars can be saved/restored.
\item Fringe recording: $\approx 5\times1$min. An observation block is generally composed of 5 consecutive files. A single file is composed of 100 scans of $\pm40\mu$m stroke (for the slowest baselines), whose individual duration ranges from $0.3$s to more $1.5$s.
\item Internal calibration of flux-splitting ratio: $\approx 5\times 30$s. A background file of 100 scans is systematically taken after the five fringe files. A shutter sequence (one beam at a time) composed of four files of 50 scans is also optionally taken. \end{itemize}

\subsection{Integrated data flow}
Although it is a visitor-instrument, PIONIER follows the ESO data flow as much as posible. From the high level point of view, PIONIER is operated via the broker of observing blocks (BOB) that executes observing blocks (OB) fetched from ESO P2PP software. Science and calibration OBs are conveniently generated by the Aspro2 and SearchCal softwares from JMMC\footnote{http://www.jmmc.fr/aspro ; http://www.jmmc.fr/searchcal}. These tools automatically fetch the target coordinates, proper motion and magnitudes from the CDS and look for calibrators in large catalogues.

The PIONIER data reduction software (\texttt{pndrs}) runs in background during the night. The angular diameters of calibration stars are automatically recovered. Consequently the final products of the pipeline are science-ready with no real need for human intervention. This allows the data quality to be assessed and decisions are taken accordingly (improve uv-coverage by taking additional points, detect a bad calibration star).

The raw data are transferred to the PIONIER archive located at IPAG on a daily basis. The pipeline is then run offline with more careful inspection of the intermediate product. The archive finally contains both the raw FITS files and the science-ready OIFITS products.

\begin{figure}[t]
   \centering
 \includegraphics[width=0.95\textwidth]{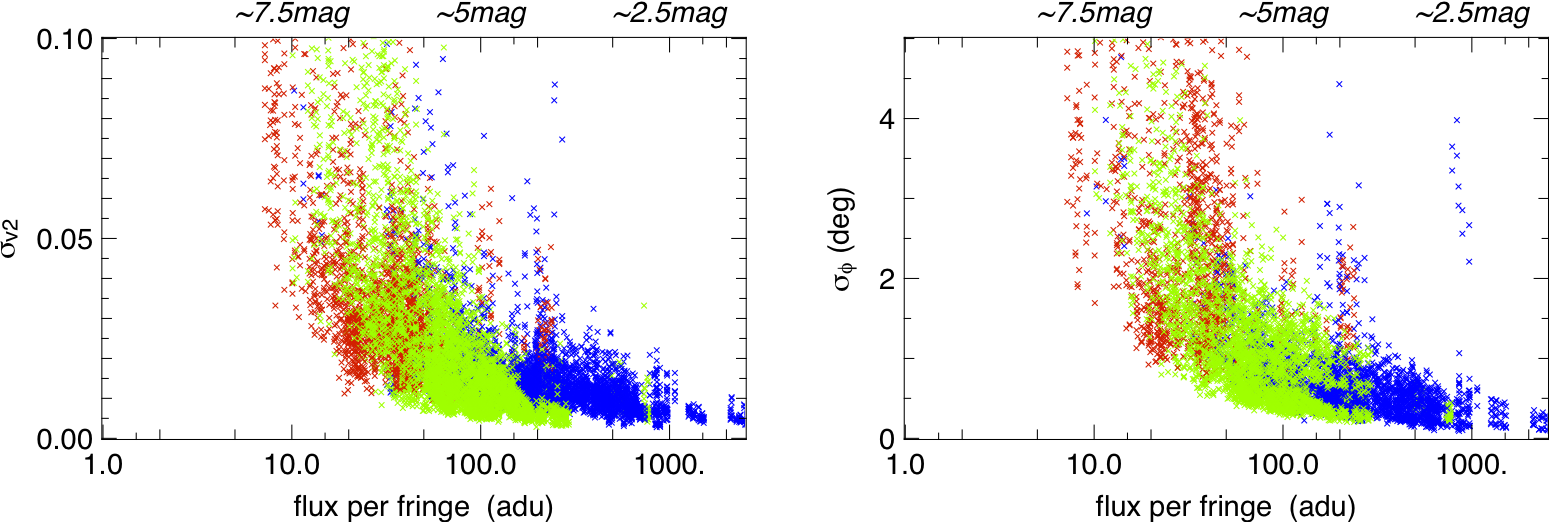} 
 \caption{Measured precision (statistical, not calibration) of the square visibilities (left) and the closure phases (right) versus the amount of flux per fringe in the scans. The colors are: blue for the dispersion over 7 spectral channels, green for dispersion over 3 spectral channels, and red for broad-band. The corresponding H-band magnitude (top) have been computed assuming median atmospheric conditions (see text). This figure represents all the PIONIER data gathered in 2012 on calibration stars only. Each point is the statistical precision obtained after averaging the 100 scans that composed a file.}
   \label{fig:stat}
 \end{figure}

\subsection{Performances}
The archive allows us to explore in a quantitative way the on-sky performances. Figure~\ref{fig:stat} summarizes the precision obtained in calibration stars for all the data gathered in 2012 so far. Exploring different parameters (such as seeing, $\tau0$, flux), we found a best correlation with the amount of flux measured in a single scan. The precision degrades dramatically at a given flux (at about 10 ADU/fringes for the broad-band mode for instance). These thresholds define empirically the sensitivity limit of the instrument in its three different modes.

The H-band magnitude corresponding to this sensitivity limit depends on the atmospheric conditions because 1) the amount of flux injected into the single-mode fiber is inversely related to the seeing, and 2) the scanning speed is adjusted depending on the coherence time. A level of 10 ADU/fringes typically corresponds to a magnitude $H=7.5$ under atmospheric condition better than the median (seeing$<1''$ and $\tau0>3$ms). Sensitivity increases to $H=8$ under very good atmospheric conditions. Accordingly, the sensitivity is reduced under poorer atmospheric conditions. These value are for the broad-band mode. The limiting magnitude is $H=6.5$ with spectral dispersion over 3 spectral channels and $H=5.5$ with dispersion over 7 channels, under median atmospheric conditions.

According to Figure~\ref{fig:stat}, the statistical precision near the limiting magnitude are typically $5\deg$ for the closure phases and $0.1$ for the square visibilities. However the statistical precision quickly reach a saturation level when increasing the level of flux. For bright targets, statistical precision are typically $0.5\deg$ for the closure phases and $0.01$ for the square visibilities. This may be interpreted as the sign of piston noise, as expected for our temporally modulated interferometric method.

\begin{figure}[!h]
   \centering
 \includegraphics[width=\textwidth]{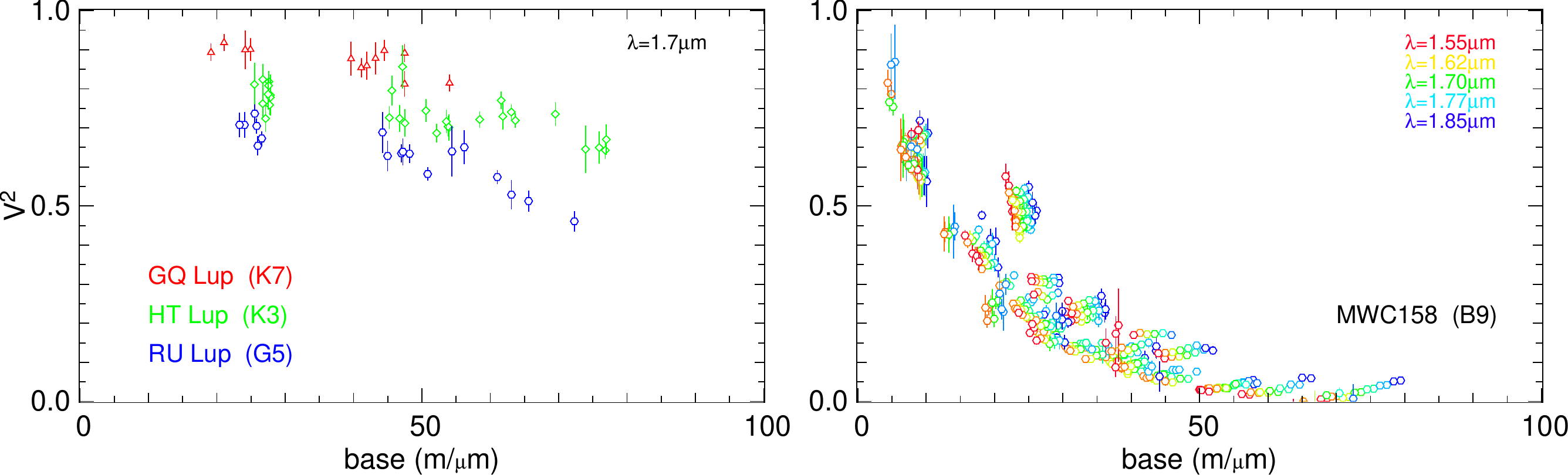}
 \caption{A representative sample of observation of Young Stellar Object with PIONIER. The square visibilities are plotted versus the spatial frequencies. Left: Observation of three T~Tauri stars in the Lupus star-forming region with the broad-band mode of PIONIER. The circum-stellar material is marginaly resolved around these faint ($H\approx7$) late type stars. It is however possible to disentangle between the compact thermal emission (slow slope in visibilities) and the more extended scattered light (visibility deficit at short baseline). Right: Intensive observation of the Herbig Be star MWC158 with the large spectral dispersion. The circum-stellar material is well resolved around this bright ($H\approx5$) early type star. The different temperature betwen the star and the disk creates a caracteristic spectral behavior of the visibilities.}
\label{fig:YSO}\vspace{2cm}
 \includegraphics[width=0.9\textwidth]{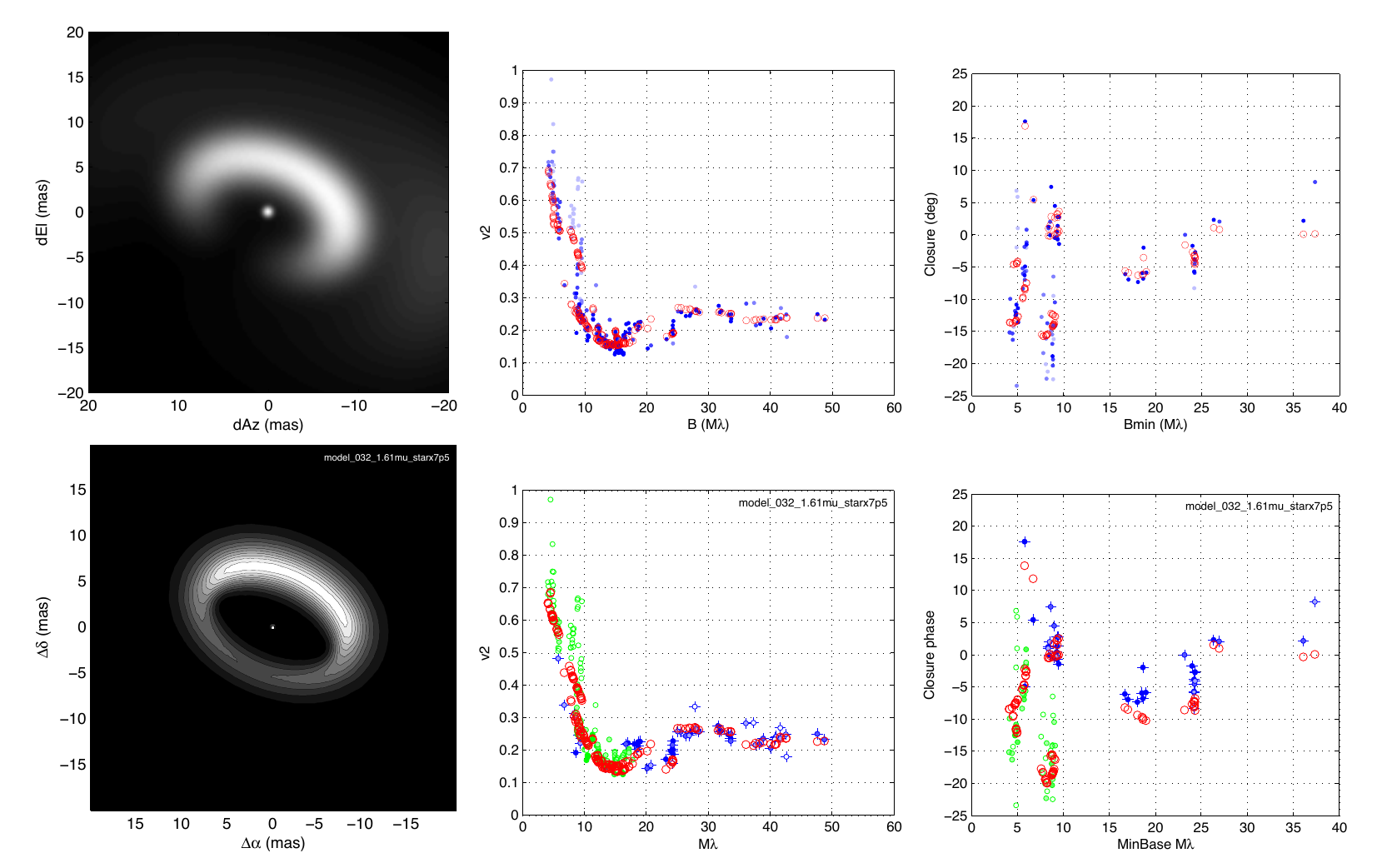}
\caption{Models versus PIONIER data for the Herbig Be object HD45677. Top: parametric modeling with a star plus a modulated disk. Below: full physics (radiative transfer, thermal and hydrostatic equilibrium) model by ProDiMo. Left to right: images of the respective models; model visibilities and phase closures (red circles) versus data (blue/green dots). For the ProDiMo model, the flux ratio between the star and the
circum-stellar material has been altered to achieve the best fit.}
\label{fig:hd45677}
 \end{figure}

\section{STATUS OF OBSERVING PROGRAMS}
\label{sec:science}

\subsection{Overall success rate at OPC}
As a visitor instrument, PIONIER has no privileged access to ESO observing time. The team competes with other VLT(I) mainstream instruments for time allocation. Proposals are submitted each semester for scientific evaluation by ESO's OPC. The average pressure on the ATs ranges from 2 to 4, with a strong peak from January to May when the galactic plane is observable. The numbers of accepted/requested nights per 6-month periods were: 24/32 nights in P87, 19/50 in P88, 30/44 in P89 and more than 50 requested night in P90 (OPC answers not known as of this writing). Additionally, a large program of 20 nights aiming at measuring the fraction of interferometric binaries among O stars is scheduled for the periods P89/90. We submitted another large program for P90/91 to study a large sample of massive young stellar objects.

\subsection{Young Stellar Objects}
One of the central goals of PIONIER is to survey a wide sample of pre-main sequence stars with different range of masses and evolutionary status. Its current limiting magnitude has permitted for the first time the observation of T~Tauris with the ATs. More than 50 pre-main sequence objects have been monitored so-far. It represents the biggest sample ever observed by an interferometric instrument. The spectral types spanned go from Be2 to K7 which offers a unique opportunity to study the dependence of the inner proto-planetary disk emission properties with the central star's one.

Figure \ref{fig:YSO} shows different visibility curves as a function of spectral type which reveals the diversity of environments. The left plot shows a sample of marginally resolved T~Tauri in the Lupus star-forming region while the right plot shows the Herbig Be star MWC~158. From the slope of the curve we confirm the correlation between the emission size and the central star spectral type (therefore luminosity) which is attributed to a putative inner dusty sublimating rim. The different temperature between the rim and the stars generates a specific comma-like structure in the visibilities (color coded in right figure \ref{fig:YSO}).

Additionally we reveal very often the presence of an extended emission that escapes our spatial frequency sampling and would require dedicated short-baselines observations. The origin of such an emission is debatable but is probably linked to light scattered by the disk at larger distances than the rim. Moreover we exploit the closure phases to constrain the emission asymmetry and relate it to the emission process itself (asymmetric scattering, radiative transfer effect) or the presence of a companion. Top right plot in figure~\ref{fig:hd45677} shows a nice example of non-zero closure phase which we attribute to the inner rim emitting asymmetry.

The data analysis methodology employed combines aperture synthesis when the amount of data and spatial extension permits it (see e.g right part of Figure~\ref{fig:YSO} and Kluska et al. same conference) with parametric model fitting (see top plots of Figure~\ref{fig:hd45677}). On the best studied targets (with e.g additional constraints from the SED, Herschel, Spitzer) we carry a multi-technique modeling using the state of the art radiative transfer code PRODIMO\cite{Thi:2011} and MCFOST\cite{Pinte:2006} (see e.g bottom of Figure~\ref{fig:hd45677}).

\subsection{Debris disks, precision visibilities}
\begin{figure}[t]
   \centering
 \includegraphics[height=6.8cm]{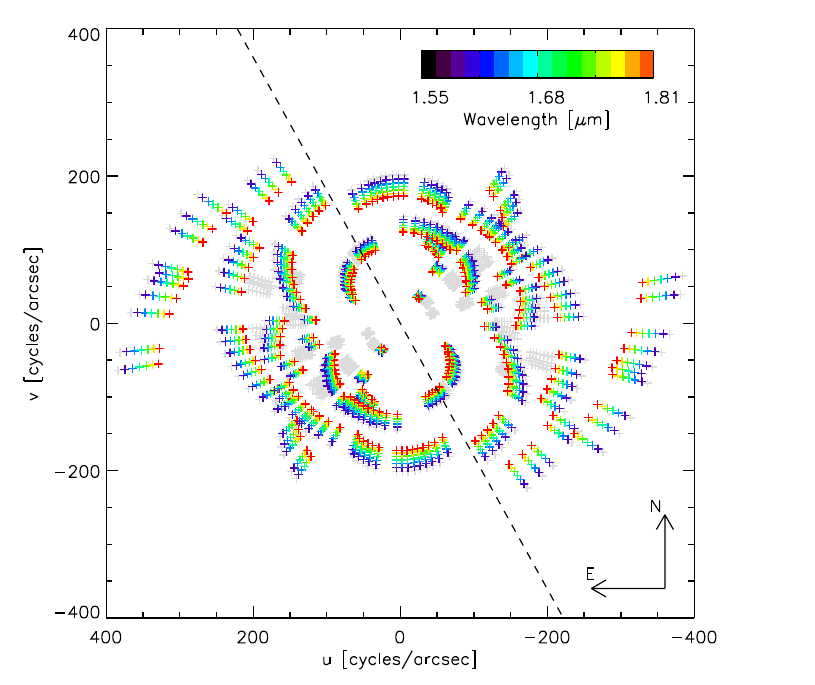}
 \includegraphics[height=6.7cm]{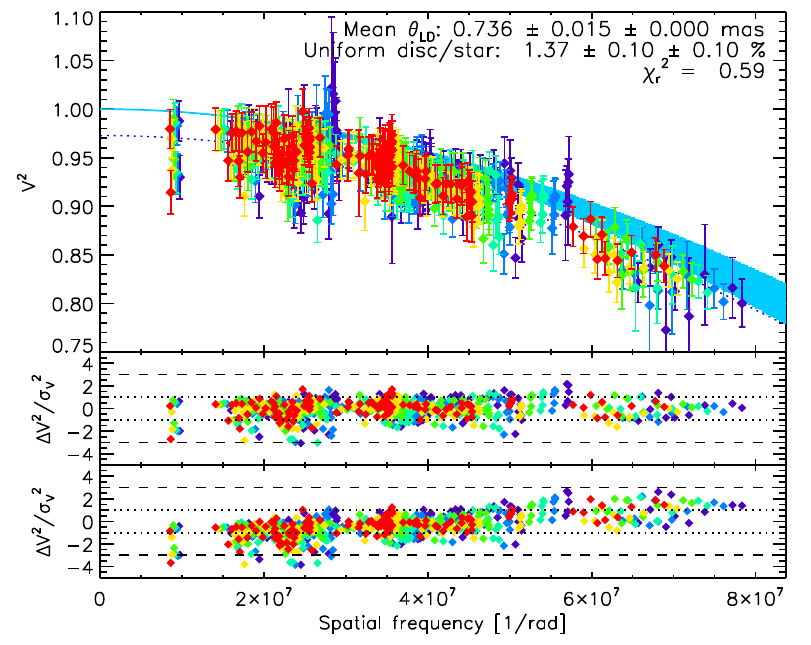} 
 \caption{Detection of a small visibility deficit ($1.37\%\pm0.2\%$) around the nearby young star $\beta$~Pic. The uv-plane observed in 3 nights is represented on the left. The measured square visibilities are plotted versus spatial frequencies on the right. Residuals are shown for the best-fit model with a Uniform Disk + visibility deficit (middle) and with a Uniform Disk only (bottom).}
   \label{fig:betaPic}
 \end{figure}

Our goal is to study the hot dust content during the late phases of terrestrial planet formation, revealed as a resolved excess flux of about 1\% (see for instance Absil et al.\cite{Absil:2008sep} and Figure~\ref{fig:betaPic}). The strategy is to survey an unbiased magnitude-limited all-sky sample of 100 stars, evenly spread between spectral types A, F and G-K-(early M). Approximately 20 targets have been already observed up to now and the first results are very encouraging.

The challenge is to obtain an accuracy of 1\% or less on the calibrated visibility. The most important observational parameter is a high scanning speed, which forces us to focus on bright ($H<5$) stars with the small dispersion only in order to reduce the number of pixels to be read. A video of beautiful fringes recorded during a great night can be found at http://www.youtube.com/watch?v=IOGtC7rkSCU

\begin{figure}[t]
   \centering
 \includegraphics[width=0.95\textwidth]{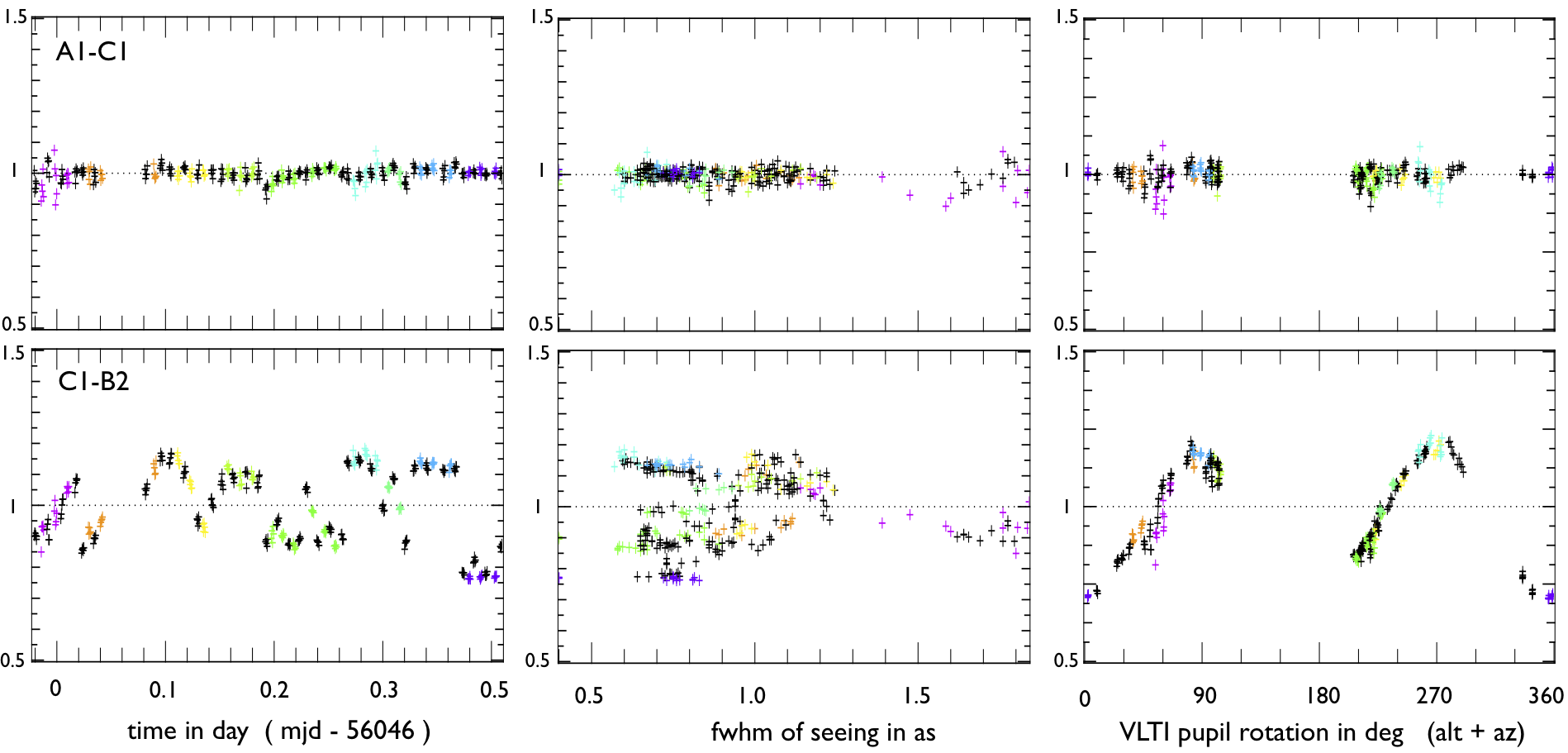} 
 \caption{Example of the transfer function of a good-quality night dedicated to precision visibilities (2012-04-29). The square visibilities for two baselines (top A1-C1 and bottom C1-B2) are plotted versus different parameters, namely: time (left), seeing (middle) and pupil rotation angle (right). Colors are for different stars (all calibration stars, almost unresolved). Data have been divided by the average transfer function of the night. True average instrumental visibility is of the order of $V^2\approx0.75$.}
   \label{fig:tf}
 \end{figure}

With this mode, the transfer function is robust against seeing fluctuations (Figure~\ref{fig:tf}, top) and other parameters such as the coherence time or the flux of the observed target. However we discover that the transfer function is highly correlated to the pupil rotation angle inside the VLTI mirror train. This correlation is negligible on some baselines, while very strong on others (compare Figure~\ref{fig:tf} top and bottom). They are not the same baselines every night, neither the same PIONIER inputs. Our primary suspect is a nasty polarisation cross-talk effect between VLTI and PIONIER. This should be further investigated. In the meantime we will observe calibration stars very close on sky of each science target, and observe calibration stars all over the sky. Both can be achieved during the night because we survey nearby stars that are distributed randomly in the sky.

Precision visibilities could also be used to measure photospheric diameters. There is no such program yet with PIONIER, arguably because of the limited length of the VLTI baselines (130m).

\subsection{Binaries and faint compagnons}
Long baseline interferometric observations naturally complement the radial velocity and adaptive optics surveys to discover and characterize binary systems. For instance, radial velocity measurements on massive stars are challenged by the lack of spectral lines and their intrinsic broadening. On active stars, the intrinsic radial velocity jitter may easily hide the signal of a faint companion. Moreover, compared with adaptive optics, the access to smaller separations and therefore shorter periods increases the possibility of determining the dynamical masses of the components.

We have initiated several surveys from massive to low-mass stars, including young stars and stars in nearby moving groups. Deep integrations on a few selected targets permit assessing what is the best possible dynamic range achievable. It is around 1:300 around bright stars such as $\beta$~Pic, Fomalhaut or $\tau$~Ceti.\cite{Absil:2011} The dynamic is typically lower around fainter stars such as TW~Hya (1:20) or T~Cha (1:100).

Note that calibration stars sometimes popup as interferometric binaries with nearly equal flux ratio. This is all the more true for spectral types A and earlier. Indeed, these binaries are hard to detect from the analysis of the spectral energy distributions as the two components have similar temperature. As of June 2012, PIONIER reported 12 bad calibrators. According to the dedicated webpage of JMMC\footnote{JMMC badcal:  http://apps.jmmc.fr/badcal}, it is the most efficient instrument to do so.

\section{LESSON LEARNED}
\label{sec:lessons}

\subsection{Performances}

\subsubsection{Usefulness of the small spectral dispersion} PIONIER features a mode with an unusually small spectral dispersion (3 channels across the H-band). Although this was not really expected, this mode appears to be a very good compromise. For Young Stellar Object, it provides valuable diagnosis to disentangle the hot photosphere from the cool environment, without sacrificing too much of the sensitivity. When searching for binaries, it enlarges the usable field-of-view by reducing bandwidth smearing up to about $50$mas, just bridging the gap with classical AO-assisted imaging. This mode is also very well suited to precision visibilities, where it mitigates chromatic issues without reducing too much the frame-rate (critical to achieve high-precision calibration). Nevertheless the larger spectral dispersion of 7 channels across the H-band is still mandatory for objects with more complexe chromatic (Mira stars) or spatial behaviors (supergiants surfaces, multiples systems).

\subsubsection{Sensitive interferometry is within reach} By reaching $H=8$ on the ATs and $H=11$ with the UTs (under good weather conditions), PIONIER defines a new standard for sensitive imaging in the near-infrared. Still, the numerical aperture of the component outputs is known to not match the aperture defined by the cold stop in the detector dewar (an heritage of the old IOTA design, which could not be improved in the time available). A gain of about 20\% in transmission is expected with an optimized design. Another breakthrough is the arrival of the new-generation low noise infrared camera (see Section~\ref{sec:future}). Considering all these improvements in an hypothetical future PIONIER-like instrument, we believe that fringe detection and coherencing should be feasible up to $H=9$ on the ATs and $H=11.5$ on the UTs.

\subsection{Imaging capabilities}

\subsubsection{The VLTI baselines}
Our expertise with PIONIER confirms that a really model-independent image requires to observe several hours with three configurations of four ATs (except very simple targets such as binaries that can be mapped with a single configuration). The offered configurations for P90 delivers a sufficiently dense and homogeneous uv-plane. As much as possible, targets with declination below $-40$deg should be favored when preparing an imaging project because they benefit from an exquisite super-synthesis effect from VLTI (see the uv-plane of Figure~\ref{fig:betaPic}).

The maximum length of the VLTI baselines (130m currently offered, possibility for a 200m baseline) is a bit short for YSO. Only an handful of Herbig Be stars could be properly imaged. The others can only be studied with parametric modeling as the spatial resolution is not sufficient to resolve the inner rim. This is even more true for Herbig A stars and T~Tauri stars.

In conclusion, imaging proposals at VLTI should be biased to optimize the image quality, even if the finally selected target (large and located toward south) is not exactly the one that would have been choose based on purely astrophysical criteria. We hope the OPC could understand this strategy. Finally, we note that OPC seems reluctant to give enough time for imaging.  One possibility would be to identify a few key targets in the range 20h$\to$6h of Right Ascension, where the competition for allocation time is not as strong.

\subsubsection{Mira stars, not so easy}
Mira stars seem a perfect case: they are large, bright, and can be found at southern declination. But they accumulate two difficulties. The chromatic behavior introduced by the molecular layers prevents from using the spectral dispersion to improve the uv-plane (each channel see a different image). Even worse: they are highly variable. The three AT configurations should be obtained within less than two weeks. If a single night is lost, the entire dataset becomes useless for imaging. Service-mode or better short-term re-scheduling of time allocation is a way to improve (another solution is to recombine more telescopes simultaneously).

\subsubsection{Stellar surfaces, an unexploited niche}
Imaging complex features at the surface of giant and super-giant stars is a key-science for optical interferometry. Routine imaging of bright and well resolved targets is largely within reach with PIONIER and VLTI. However it requires an additional effort in commissioning, mainly: 1) verifying the detector mode DOUBLE, 2) investigating the accuracy and stability of the spectral calibration, and 3) checking the quality of the VLTI baseline model. So far these aspects were put a low priority as the team focused into developing an optimal strategy to observe faint and/or barely resolved targets (the primary scientific interests of IPAG are YSO, debris disks and faint companions).

\subsection{Building a fast-track instrument}
Most of the shortcuts are related to the lack of time to gain sufficient knowledge on the instrument before shipping (see section 2.3).  On the other side, we identify three key-ingredients that immensely benefit to the PIONIER project\footnote{The PI of the instrument would like to add another point: the professionalism and high level of expertise of the technical staff that designed and built PIONIER, most of them being co-authors of this article.}:\begin{itemize}
\item Members of the PIONIER team had an in-deepth knowledge of ESO environment, programming standard, and Paranal operations, complemented by a practical experience of interferometric instrumentation.
\item The project benefited from the easy interactions and mutual trust between ESO and the PIONIER team, and from the very good support from Paranal and Garching from the design to the operation of the instrument\footnote{Among other contribution: extend VLT software to support a novel solution for instrument control, help in designing the cryogenic circuits, provide a system to absorb vibrations from the electronic cabinet.}. ESO carried out the transport of the instrument from Europe to Paranal.
\item At each stage of the project, we focused on hardware components, instrumental concepts and operational strategies already demonstrated on-sky or for which the team already had previous experience, although they were known to be non-optimal. A well implemented sub-optimal concept is generally better than a clever idea poorly implemented.
\end{itemize}

\subsection{Observing at VLTI with a visitor-instrument}
\subsubsection{Interest of a simple, reliable instrument} The reliability of VLTI operations considerably improved during the last years. When using the four Auxiliary Telescopes the majority of the downtime now comes from weather losses (wind, humidity, clouds, bad seeing) rather than technical losses. However operations of the complexe VLTI machinerie are affected by numerous, small ``operational glitches'', such as aborted preset or guiding failure. In this context, the capability  to quickly recover brought by a simple and reactive instrument is highly valuable. Based on similar considerations, a simple and reactive instrument has a higher probability to efficiently use a short time window of good seeing.

\subsubsection{Difficulties in run scheduling} PIONIER runs are scheduled as any other visitor runs at VLTI: they are spread over the entire 6-month period. This large number of short runs requires frequent flight to Chile, which is inconvenient. We see two possibilities to improve: 1) a member of the PIONIER team communicates with ESO Observing Programmes Office at the time of scheduling, or 2) we submit only few, general purpose Large Programs instead of numerous smaller proposals focused on specific targets (as of now).

\subsubsection{Interest of real-time optimisation} The PIONIER observer has full freedom to adapt the observing strategy in real-time. All instrumental parameters can be adjusted at night, new ideas can be tested, optimization (or bug-fix) of the software can be implemented. We found that the pros of such an agressive strategy (gathering optimal data) largely overcompensate the cons (possibility of mistakes and time losses).

\subsubsection{Training new students} Finally, PIONIER is a great opportunity for students to learn practical interferometry, including aligning the instrument, optimizing the observations, and analyzing the data. So far, more than five new students or postdoc have been trained for operating PIONIER in standalone. We note that the second generation instruments MATISSE and GRAVITY gather the expertise of former students trained in GI2T, IOTA and CHARA.

\section{THE FUTURE OF PIONIER}
\label{sec:future}

The continuation of PIONIER operation at Paranal is discussed and renewed on a yearly basis. Current agreement lasts until October 2013. Until this date, the instrument is expected to be operated in H-band and occasionally in K-band within the same framework as presented in this paper.

In the medium term, the arrival of a new generation of infrared detectors will provide detector noise as low as 2.5e$^-$ at a frame-rate higher than 1 kHz in destructive mode\cite{Finger:2010}. IPAG is involved in a large collaboration aiming at developing such a detector, called RAPID. First working prototypes are expected end of 2012 and will be tested at IPAG. A dedicated test bench recreating hardware and software conditions similar to the VLTI laboratory has been developed. It should allow an easy relocation of the detector into the PIONIER instrument. The chance of this project to succeed partially depends on how long PIONIER will remain in Paranal, as the current schedule is very tight. Installed in PIONIER, RAPID would increase the sensitivity by 1.5mag, improve the read-out speed and solve saturation problems.

Finally, with the arrival of the GRAVITY instrument, PIONIER will have to be dismounted from its current location in the VLTI laboratory. Several options to continue the operations have been envisioned by different actors, although none of them have been effectively discussed in details (shipping PIONIER in another interferometer, upgrading PIONIER to replace AMBER or FINITO).

\acknowledgments 

PIONIER is funded by the Universit\'e Joseph Fourier (UJF, Grenoble) through its Poles TUNES and SMING and the vice-president of research, the Institut de Plan\'etologie et d'Astrophysique de Grenoble, the ``Agence Nationale pour la Recherche'' with the programs ANR EXOZODI and ANR POLCA, and the Institut National des Science de l'Univers (INSU) with the programs ``Programme National de Physique Stellaire'' and ``Programme National de Plan\'etologie''. The integrated optics beam combiner is the result of a collaboration between IPAG and CEA-LETI based on CNES R\&T funding. The authors want to warmly thank all the people involved in the VLTI project. This work is based on observations made with the ESO telescopes. It made use of the Smithsonian/NASA Astrophysics Data System (ADS) and of the Centre de Donnees astronomiques de Strasbourg (CDS). All calculations and graphics were performed with the freeware \texttt{Yorick}.


\bibliography{/Volumes/Datas/lebouquj/Biblio/BibTex/allNew}   
\bibliographystyle{spiebib}   

\end{document}